\documentclass[%
 sor,
aip,
 jor,
 amsmath,amssymb,
 reprint,%
author-numerical,%
]{revtex4-2}

\usepackage{graphicx}
\usepackage{dcolumn}
\usepackage{bm}
\usepackage{xcolor}
\usepackage{wrapfig}
\usepackage{lipsum}  

\begin{document}

\preprint{AIP/123-QED}

\title[Forced ghost cycles and slow-fast systems]{Ghost cycles exhibit increased entrainment and richer dynamics in response to external forcing compared to slow-fast systems}

\author{Daniel Koch}
 \email{daniel.koch@mpinb.mpg.de}
\author{Aneta Koseska}
 \email{aneta.koseska@mpinb.mpg.de}
\affiliation{ 
Cellular Computations and Learning Group, Max Planck Institute for Neurobiology of Behavior - caesar, Ludwig-Erhard-Allee 2,
53175 Bonn, Germany.
}%

\date{\today}

\begin{abstract}

Many natural, living and engineered systems display oscillations that are characterized by multiple timescales. Typically, such systems are described as slow-fast systems, where the slow dynamics result from a hyperbolic slow manifold that guides the movement of the system trajectories. Recently, we have provided an alternative description in which the slow dynamics result from a non-hyperbolic and Lyapunov-unstable attracting sets from connected dynamical ghosts that form a closed orbit (termed ghost cycles). 
Here we investigate the response properties of both type of systems to external forcing.
Using the classical Van-der-Pol oscillator and two modified versions of this model that correspond to a 1-ghost and a 2-ghost cycle, respectively, we find that ghost cycles are characterized by significant increase especially in the 1:1 entrainment regions as demonstrated by the corresponding Arnold tongues and exhibit richer dynamics (bursting, chaos) in contrast to the classical slow-fast system. Phase plane analysis reveals that these features result from the continuous remodeling of the attractor landscape of the ghost cycles models characteristic for non-autonomous systems, whereas the attractor landscape of the corresponding slow-fast system remains qualitatively unaltered. We propose that systems containing ghost cycles display increased flexibility and responsiveness to continuous environmental changes.

\end{abstract}

\maketitle

\begin{quotation}
Oscillations in natural systems often exhibit multiple time-scales, characterized by repeated switching between slow and fast motions. Such phenomena are typically modeled as slow-fast systems, whose dynamics formally corresponds to a type of stable limit cycle oscillators. We have recently proposed that similar dynamics can also arise by connecting unstable objects called ghost attractors forming a cycle, such that there is slow dynamics due to the ghosts, with fast switching between them. We show here that ghost cycles exhibit a higher flexibility in their response to periodic external inputs, allowing them to be more easily entrained in to a broad range of frequencies of the external input, but also to exhibit complex dynamics such as bursting and chaos. 
Our analyses show that this flexibility results from an organization at criticality, such that the ghost cycles can exploit different dynamical regimes present in the system 
under the influence of time-varying external inputs. We validate the results for different models and discuss the implications of these findings for information processing tasks in biological systems.

\end{quotation}

\section{\label{sec:level1} Introduction}

Many natural, living and engineered systems display behavior that is periodic or oscillatory. Examples include respiration, circadian clocks \cite{Roenneberg_2016}, pulsatile hormone secretion \cite{Brabant_1992}, neural and cardiac rhythms \cite{Glass_2001,Izhikevich_2006}, chemical and biochemical reactions \cite{Noyes_1974,Goldbeter_2013}, population cycles of predator-prey types \cite{VOLTERRA_1926,Elton_1942,Blasius_2019}, insect-outbreaks \cite{Liebhold_2004}, climate phenomena \cite{Wang_2016,Dijkstra_2005,Vettoretti_2018,NGIPCM_2004}, lasers \cite{laserbook}, to name but a few. One of the most common type of periodic behaviors are relaxation oscillations, characterized by repeated switching between slow and fast motions. These processes are generally modeled by singularly perturbed ordinary differential equations \cite{Kuehn_2015,Bertram_2017} of the form

 \begin{equation}
    \dot{x} = f(x,y,\lambda), \quad 
    \dot{y} = \varepsilon g(x,y,\lambda),
     \label{eq:slow-fast}
 \end{equation}

\noindent where $(x,y)\in \mathbb{R}^n \cdot \mathbb{R}^m$, and $0< \varepsilon \ll 1$ is a small parameter that determines the time-scale separation. Considering the singular limit as $\varepsilon\rightarrow0$, the dynamics of such slow-fast systems is characterized by the critical manifold, defined as $C=\{(x,y)\in \mathbb{R}^n \cdot \mathbb{R}^m \| f(x, y, \lambda, 0)=0\}$. A normally hyperbolic manifold $S \subset C$ that fulfills the criteria of Fenichel theory\cite{Kuehn_2015,Fenichel_1971} is denoted as a slow manifold and is typically used to describe the dynamical properties of the system.

There is a growing recognition, however, that a multitude of naturally occurring processes are characterized by quasi-stable dynamics with fast switching between them, that cannot be fully described by the asymptotic dynamics captured by slow-fast systems. This includes neuronal activity dynamics during  rest and behavioral or sensory tasks \cite{beim_Graben_2019,Kato_2015,Woo_2023}, coral to macroalgae dominance in coral reefs\cite{Bieg_2024}, long-transients in other ecological systems \cite{Hastings_2018} and embryonic development \cite{Sanchez_2022,Meeuse_2020}, among others. Thus, complementary to the idea of attractors governing system's dynamics, it has been proposed that long transients can emerge due to the systems dynamics lingering near a dynamic saddle or is guided through heteroclinic channels / cycles  \cite{Rabinovich_2001,Afraimovich_2004,Ashwin_2013}. Complementing this description, we have recently suggested that such sequential quasi-stable dynamics can additionally emerge when the system's dynamics is organized on scaffolds of ghost structures, particularly cyclic organization of ghosts \cite{Koch_2023}. This is dynamically equivalent to an organization in a vicinity of one or multiple co-occurring saddle node on invariant cycle (SNIC) bifurcations. In this case, the dynamics of the system is organized around a set that is non-hyperbolic, and the flow in phase space is guided by a phase-space structure that is not-stable in Lyapunov sense \cite{Gorban_2004,Gorban_2013}.
Thus, experimental observations of processes that have quasi-stable dynamics with fast switching between seemingly stable periods can emerge from different dynamical mechanisms in autonomous systems that are often indistinguishable based solely on the time-series characteristics. 

Natural systems in particular, however, do not operate in isolation, but are rather subjected to external signals that vary over space and time. This is true for cellular systems, where the signals are produced by the neighbouring cells or tissues, neuronal systems that are continuously exposed to non-stationary sensory signals as it is for ecological systems that are influenced by the weather, climate or human-induced forces. The system's response to external forcing is therefore likely determined by the mechanism through which the autonomous dynamics arises. 
To investigate whether attractor-governed slow-fast systems and ghost cycles characterized by non-asymptotic transients dynamics differ with respect to their response properties when driven by external signals, we study here the entrainment and response dynamic characteristics of the Van der Pol oscillator (VdP) as a prototypical slow-fast system and two modified Van der Pol systems which exhibit one-ghost (VdP$_\text{1g}$) and two-ghost (VdP$_\text{2g}$) cycles, respectively. We show that under low-amplitude periodic forcing, ghost cycles not only show a strong increase in the Arnold tongue area, i.e. increased entrainability, but also qualitatively new behaviors such as bursting and chaos. These features can be understood from the fact that ghost cycles emerge at criticality, allowing the corresponding non-autonomous system to exploit different dynamical regimes as a consequence of a continuous change of the attractor landscape.

\section{\label{sec:results} Results}

\begin{figure}[h]
\includegraphics[width=\textwidth]{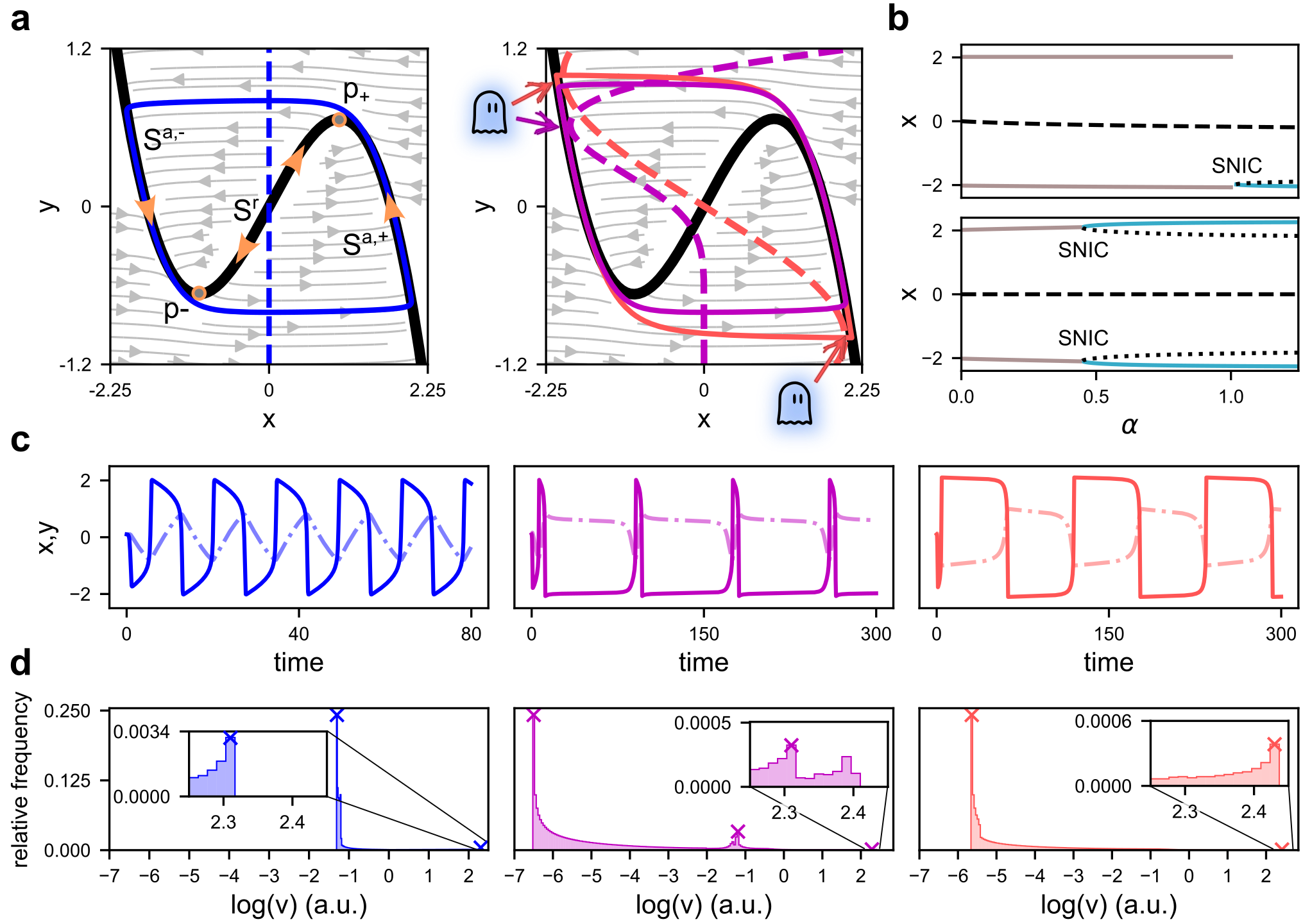}
\caption{Emergence of ghost cycles in a modified VdP oscillator. (a), left panel: phase space organization of the original VdP model. Black line: x-nullcline, dashed blue line:  y-nullcline, continuous blue line: system's trajectory. The critical manifold of the VdP corresponds to the x-nullcline and consists of three branches $S^{a,+},S^r,S^{a,-}$, the slow flow along which is indicated by orange arrowheads. Right panel: the y-nullclines of VdP$_\text{1g}$ (purple dashed line) and VdP$_\text{2g}$ (red dashed line) models for $\alpha = 1.01373$ and $\alpha = 0.44431$, respectively, leading to the emergence of ghost attractors (arrows). Trajectories are shown in continuous purple (VdP$_\text{1g}$) and red (VdP$_\text{2g}$) lines. (b) Bifurcation analysis of the  VdP$_\text{1g}$ (top) and  VdP$_\text{2g}$ (bottom). Dashed black line: unstable spiral; grey line: stable limit cycle; dotted black line: saddle; teal line: stable fixed point. SNIC bifurcations at $\alpha = 1.01873$ and $\alpha = 0.44931$, respectively. (c) Simulated time-series of x- (continuous line) and y-variables (dashed/dotted line) of the VdP (left), VdP$_\text{1g}$ (middle) and VdP$_\text{2g}$ (right) systems. (d) Histograms of the velocity distributions along the systems' trajectories. Indicated peaks are taken as representative time-scales of the systems.}
\label{fig:fig1}
\end{figure}

\subsection{Introducing ghosts in the Van der Pol system: effect on oscillations and time-scales}

Let us consider the Van der Pol (VdP) model \cite{van_der_Pol_1928}, a classical slow-fast system given by:
\begin{align}\label{eq:Vdp}
    \nonumber \dot{x} &= \frac{1}{\varepsilon}(x-\frac{x^3}{3}-y)  \\ 
    \dot{y} &= \varepsilon x  + A\ \sin(\omega t),
\end{align}

\noindent where $\varepsilon$ is the time-scale separation parameter, and $A$ and $\omega$ are the amplitude and the frequency of external periodic forcing, respectively. In the autonomous case ($A=0$) and for $\varepsilon=\frac{1}{7}$, the VdP oscillator exhibits a pronounced separation of the slow and fast time-scales and oscillates with an intrinsic frequency $\omega_0$. The critical manifold in this system is the cubic curve $y=x^3/3-x$, which is the x-nullcline of the system, whereas the y-nullcline is defined by the line $x=0$ (Figure \ref{fig:fig1}a, left). The critical manifold is normally hyperbolic away from the local maximum (p$_\text{+}$) and minimum (p$_\text{-}$) of the cubic, and these fold points decompose the critical manifold into two slow and attracting (S$^\text{a,-}$, S$^\text{a,+}$) and one repelling branch (S$^\text{r}$). Thus, a trajectory starting on the slow manifold moves along the slow branch before rapidly switching at the fold point to the second slow branch. This leads to temporal dynamics (Figure \ref{fig:fig1}c, left) characterized by two distinct trajectory velocities indicative of a slow-fast system, which is further reflected in the bimodal histogram of the velocity distribution (Figure \ref{fig:fig1}d, left).

Ghost cycles, on the other hand, do not rely on a hyperbolic slow manifold. In fact, the non-hyperbolicity of ghost dynamics results from an eigenvalue gradient from negative to positive including zero eigenvalues within in the attracting set \cite{Koch_2023}. Rather, ghost dynamics  emerges when the nullclines of the system are close to each other but not intersecting, i.e. when the system is close to a SNIC bifurcation, as noted above. 
Thus, by manipulating the shape of the $y-$nullcline, it is possible to introduce ghost dynamics in the VdP system, in addition to the slow-fast dynamics. Consider the following modifications:
\\ \\
\noindent (VdP$_{\textnormal{1g}}$ system)
\begin{align}\label{eq:Vdp1g}
    \nonumber \dot{x} &= \frac{1}{\varepsilon}(x-\frac{x^3}{3}-y) \\ 
    \dot{y} &= \varepsilon x + \alpha \left(y+0.7-\frac{(y+0.7)^3}{3}\right)\left(\frac{1+\tanh(y+0.7)}{2}\right)^{10} + A\ \sin(\omega t).
\end{align}

\noindent and  (VdP$_{\textnormal{2g}}$ system)
\begin{align}\label{eq:Vdp2g}
    \nonumber \dot{x} &= \frac{1}{\varepsilon}(x-\frac{x^3}{3}-y) \\
    \dot{y} &= \varepsilon x + \alpha y - \frac{\alpha y^3}{3} + A\ \sin(\omega t),
\end{align}

\noindent For $\alpha = 0$, both system (\ref{eq:Vdp1g}), (\ref{eq:Vdp2g}) are equivalent to the classical Van der Pol oscillator (\ref{eq:Vdp}). Setting $\alpha > 0$, however, controls the shape of the the y-nullcline such that it bends and approaches in the vicinity of the x-nullcline once (VdP$_{\textnormal{1g}}$) or twice (VdP$_{\textnormal{2g}}$), giving rise to a single or two ghosts (Figure \ref{fig:fig1}a, right). Formally, this corresponds to parametric organization before single or multiple SNIC bifurcations that destroy the limit cycle at a certain $\alpha_{crit}$ (Figure \ref{fig:fig1}b).

Introducing a single ghost in the VdP dynamics effectively does not change the shape of the oscillations. However, longer transients between the spikes are present, resulting from the system's dynamics visiting the ghost that is formed (Figure \ref{fig:fig1}c, middle). The additional slow time scale from trapping in the ghost is also visible in the velocity histograms (Figure \ref{fig:fig1}d, middle).
In the VdP$_{\textnormal{2g}}$ system, however, a two-ghost cycle emerges and defines the system's dynamics: while rapid transitions along the x-direction still occur, the time during which the trajectory is trapped in the ghost regions fully dominates the slow dynamics (Figure \ref{fig:fig1}c, right), rendering the influence of the original slow manifolds of the VdP system insignificant. This is also visible in the velocity histogram which reveals the peaks of the fast time-scale and the ghost-trapping time scale (Figure \ref{fig:fig1}d, right). 

As natural systems are inherently noisy, an important question is how the time-scale separation of a system is affected by noise. Previously, we have shown that the period of ghost cycles remains unaltered by low and intermediate noise intensity, becoming progressively more variable and decreasing for high noise intensity due to a loss of the slow time scale \cite{Koch_2023}. Here, we find the equivalent dependencies of the period for the VdP$_{\textnormal{1g}}$ and VdP$_{\textnormal{2g}}$ for increasing noise intensity, whereas the period of the VdP system remains generally unperturbed (Figure S1).

\begin{figure}[h]
\includegraphics[width=\textwidth]{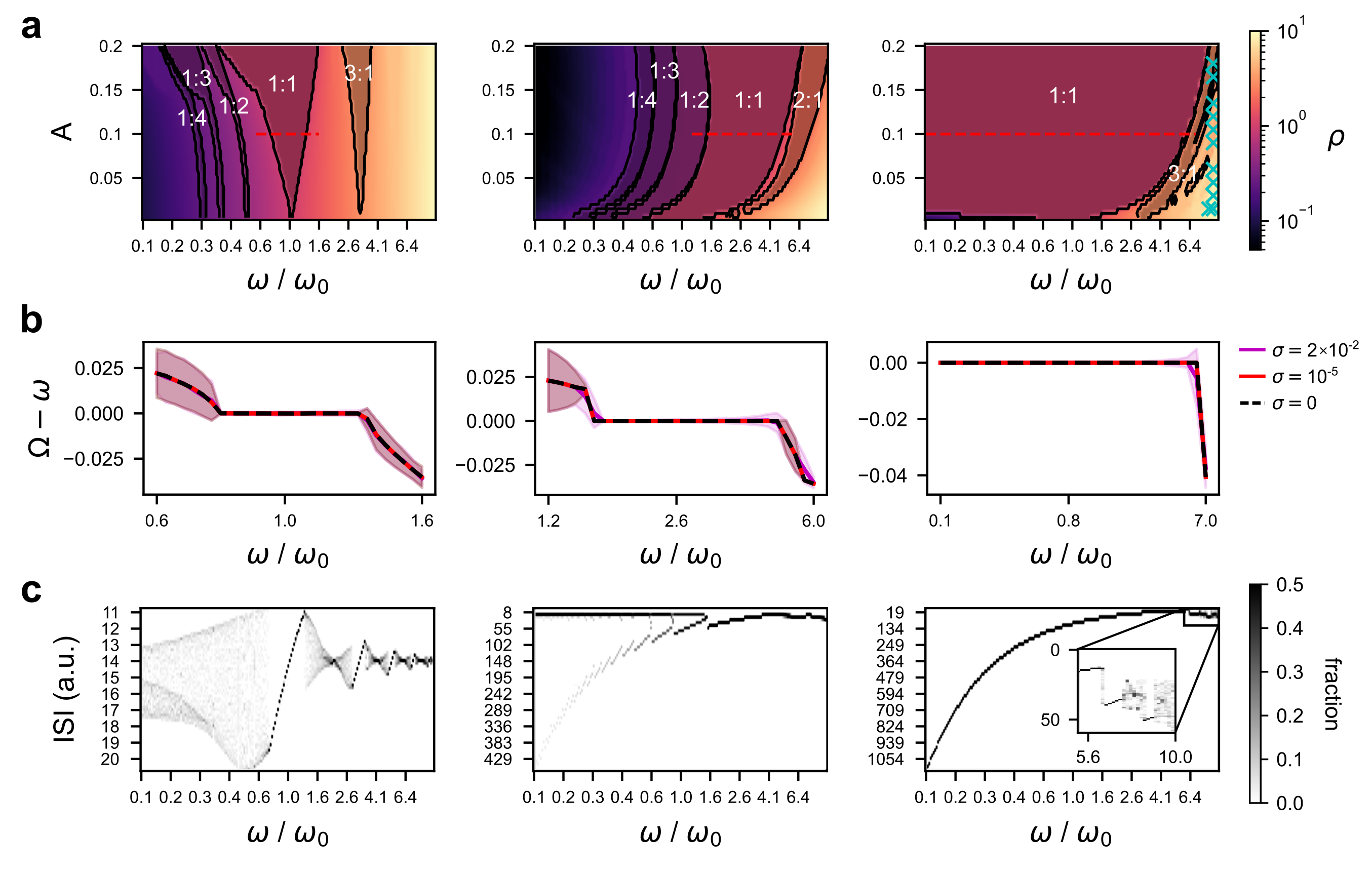}
\caption{Response to periodic forcing. (a) Arnold tongues indicating areas of n:m entrainment for the VdP (left), VdP$_\text{1g}$ (middle) and VdP$_\text{2g}$ (right) systems as a function of the amplitude $A$ and frequency ($\omega$; normalized to the frequency of the unforced system, $\omega_0$) of the forcing. Color encodes the winding number $\rho = \frac{T_{out}}{T_{in}}$; teal crosses indicate largest Lyapunov exponents $>0$. (b) Difference of the observed frequency $\Omega$ and the driving frequency $\omega$ at various noise intensities in the parameter range indicated by the dashed red lines in (a). Plateaus correspond to cross-sections through the 1:1 entrainment tongues. (c) Distribution of the interspike intervals (ISIs) as a function of the forcing frequency for the VdP (left), VdP$_\text{1g}$ (middle) and VdP$_\text{2g}$ (right) system.}
\label{fig:fig2}
\end{figure}

\subsection{Response to periodic forcing}

To investigate whether the different mechanisms through which the slow time-scale emerges in the system are also reflected in differences of how the systems respond to time-varying signals, we subjected the VdP, VdP$_{\textnormal{1g}}$ and VdP$_{\textnormal{2g}}$ systems to periodic forcing, and numerically quantified the corresponding Arnold tongues\cite{Pikovsky_2001}. 
Arnold tongues visualize the entrainment behavior of forced oscillatory systems and represent areas in the forcing amplitude-frequency space in which the system is mode-locked to the external forcing as represented by a constant winding number $\rho$ \cite{Ram_rez_vila_2018}. Typically, major entrainment areas where the forced oscillator exhibits $n$ periods for $m$ periods of external forcing ($n:m$ entrainment) are highlighted.
We focus here on low amplitude forcing ($0<A\leq 0.2$) and for each system scanned an interval of forcing frequencies ranging from $0.1$- to $10$-fold of the intrinsic frequency of the autonomous system ($\omega_0$).

The VdP system exhibits a typical response to periodic forcing, where 1:1 entertainment is most prominent and $n:m$ regions exist in narrow parameter regions (Figure \ref{fig:fig2}a, left). The single ghost introduced in VdP$_{\textnormal{1g}}$, in contrast, notably changes the response properties of the system: the Arnold tongues are much wider already at low amplitude of the external forcing compared to the classical VdP oscillator, and their width is only marginally affected as the amplitude increases (Figure \ref{fig:fig2}a, middle). For the  VdP$_{\textnormal{2g}}$ system, the difference in the system's responsiveness is even more pronounced: the majority of the two parameter plane spanned by the amplitude and frequency of the driving signal is filled by the 1:1-entrainment tongue, sided only by a narrow 3:1-tongue (Figure \ref{fig:fig2}, right). 

\begin{figure}[h]
\includegraphics[width=\textwidth]{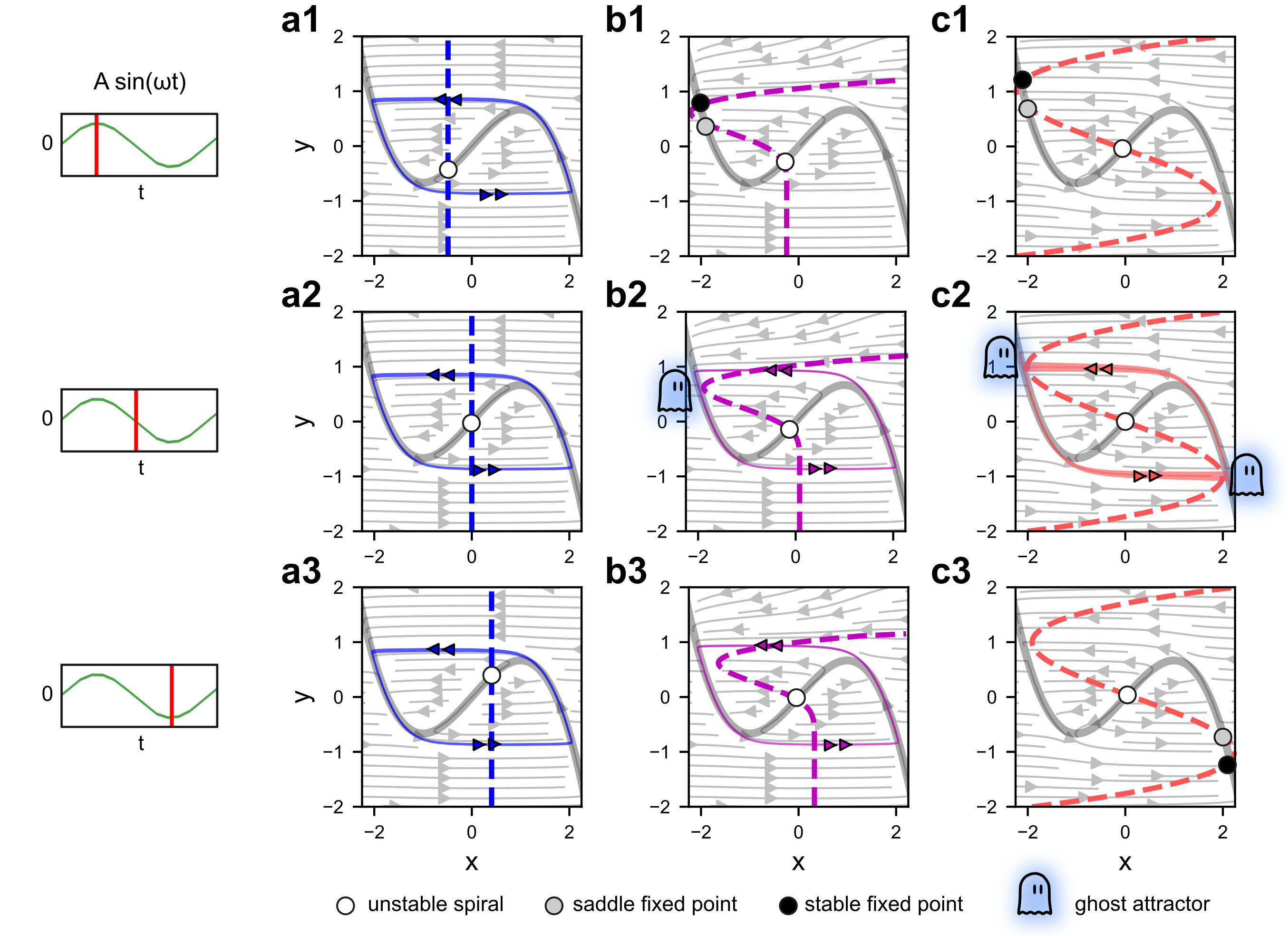}
\caption{Phase-plane analysis at different intervals of periodic forcing (left). (a1-a3) In the original VdP system, low amplitude external forcing leads to shifting of the y-nullcline back and forth along the x-axis, without inducing major qualitative changes in the attractor landscape. (b1-b3) In the VdP$_\text{1g}$ system, shifting of the y-nullcline back and forth along the x-axis leads to a periodic disappearance of the ghost attractor either via creation of a new fixed point (b1) such that the trajectory is trapped here, or via moving the nullclines further apart (b3), thereby loosing the slow time-scale from the ghost. (c1-c3) In the VdP$_\text{2g}$ system, the external forcing leads to alternate creation and destruction of fixed points, and thereby periodic disappearance of the ghost attractors. Double arrowheads denote fast transitions of the trajectory.}
\label{fig:fig3}
\end{figure}

We therefore asked what is underlying these striking differences between the VdP oscillator and its ghost cycle variants. Generally, in non‐autonomous systems or systems under the influence of time-varying inputs, the geometry (positioning, shape and size of the attractors), or topology (number or stability of the attractors) of the underlying phase space changes \cite{Verd_2014, Stanoev_2020}. Thus, we performed next a phase plane analysis at different intervals of the periodic forcing for the three distinct systems. For the VdP system, periodic forcing leads to shifting of the y-nullcline back and forth along the x-axis, without inducing major qualitative changes in the geometry or topology of the attractor landscape (Figure \ref{fig:fig3}a1-a3). Thus, the response of the forced oscillator is solely governed by the slow-fast dynamics of the autonomous systems, explaining the typical entrainment response (Supplementary Figure 2 and Video 1).

\begin{wrapfigure}{R}{0.5\textwidth}
  \begin{center}
    \includegraphics[width=0.48\textwidth]{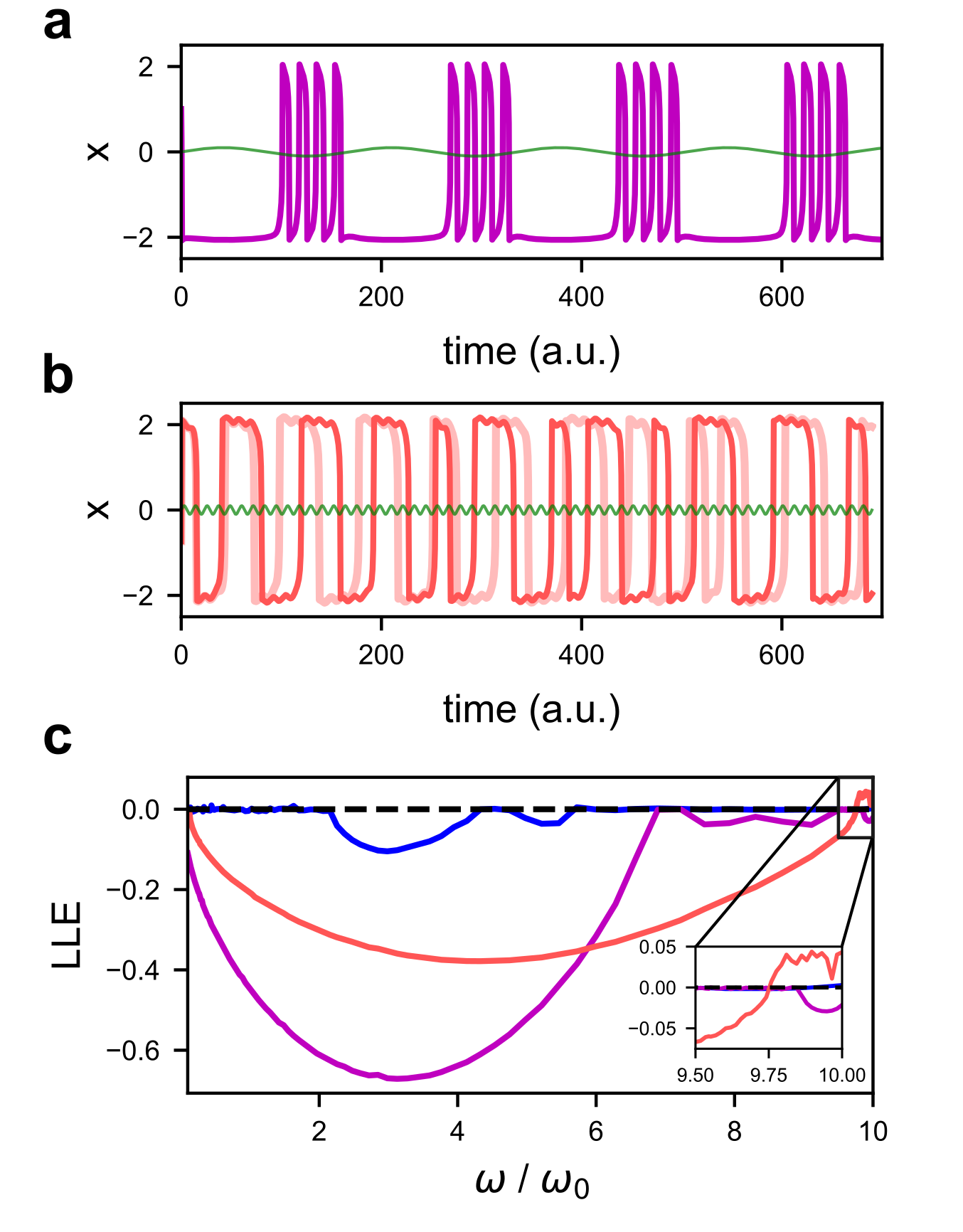}
  \end{center}
  \caption{Complex dynamics in the periodically forced ghost cycles. (a) Bursting behavior in the VdP$_\text{1g}$ system at low forcing frequencies (green), independent of amplitude (here: $A = 0.1, \omega = 0.5 \cdot \omega_0$). 
  (b) Complex periods and high sensitivity to initial conditions in the VdP$_\text{2g}$ system at high forcing frequencies (here: $A = 0.1, \omega = 10 \cdot \omega_0$). Exemplary time-series for different initial conditions are shown.
  (c) Largest Lyapunov exponent (LLE) as a function of the forcing frequency for the VdP (blue), VdP$_\text{1g}$ (purple) and VdP$_\text{2g}$ (red) system.}
  \label{fig:fig4}
\end{wrapfigure}

In the VdP$_{\textnormal{1g}}$ system, in contrast, the external forcing continuously remodels the attractor landscape. New fixed points (stable and saddle) emerge, such that the stable fixed points traps the system trajectory (Figure \ref{fig:fig3}b1), while at other times, the ghost attractor is  destroyed by moving the x- and the y-nullclines further apart (Figure \ref{fig:fig3}b3). This creates a gating mechanism in which the trajectory quickly cycles along the orbit of the  VdP$_{\textnormal{1g}}$ system when the ghost attractor is absent, but is transiently captured by the emergence of the stable fixed point at sufficient forcing strength, thus explaining the larger Arnold tongues (Supplementary Figure 3 and Video 2).

In the VdP$_{\textnormal{2g}}$ system, the periodic forcing results in a continuous alternation between the creation and destruction of stable fixed points on the diagonal of the phase space (Figure \ref{fig:fig3}c1-c3): when the input moves the y-nullcline to the left, a fixed point is created via a SNIC bifurcation, while another one on the other side of the phase space is destroyed. Moving it towards the right results to the equivalent situation in this phase-space area. This leads to a situation in which the trajectory, whenever it is released from one corner of the phase space after the destruction of the stable fixed point, is immediately captured by the stable fixed point that is generated at the other side (Supplementary Video 3). Thus, the trajectory can only move along half of the orbit at once, explaining the ultra-large range of 1:1 entrainment of the VdP$_{\textnormal{2g}}$ system (Supplementary Figure 4).

Considering the variability in the period of VdP$_{\textnormal{1g}}$ and VdP$_{\textnormal{2g}}$ for intermediate noise levels (Figure S1), we investigated next how robust the observed entrainment properties are in the presence of noise. Since Arnold tongues are not defined in the presence of noise, one can instead plot the difference between the observed and the forcing frequency as a function of the forcing frequency at a given amplitude\cite{Pikovsky_2001}. Plateaus in these plots essentially correspond to cross-sections through the Arnold tongues. Focusing on the 1:1 entrainment of the three systems at an amplitude of $A=0.1$ (cf. dashed red lines in Figure \ref{fig:fig2}a), we find that the plateau corresponding to the 1:1-tongue of the VdP system remains identical to noise-free case (Figure \ref{fig:fig2}b, left), consistent with the system's insensitivity to noise.  Similarly, even for noise levels at which high variability in the period is observed, the plateaus corresponding to the $1:1$-tongues of  VdP$_{\textnormal{1g}}$ and  VdP$_{\textnormal{2g}}$ reflect closely the characteristics of the noise-free case (Figure \ref{fig:fig3}b, middle and right), suggesting that the observed entrainment properties are robust in the presence of noise. 

Entrainment, however, is not the only difference in the behavior observed for periodic forcing of the VdP, VdP$_{\textnormal{1g}}$ and VdP$_{\textnormal{2g}}$ systems. Quantifying the distribution of the interspike-intervals (ISIs) as a function of forcing frequency at a fixed amplitude ($A=0.1$) reveals further differences in the temporal dynamics between the three systems. In case of the classical VdP system, the ISI distributions outside the entrainment bands are broad and disperse, forming patterns indicative of complex dynamics (Figure \ref{fig:fig2}c, left). For the VdP$_{\textnormal{1g}}$ system, however, the ISI distributions are bimodal at low forcing frequencies, and transit to unimodality above $\omega \approx 1.6 \cdot \omega_0$ (Figure \ref{fig:fig2}c, middle). This bimodal ISI distribution result from the emergence of bursting dynamics with trains of rapid spiking followed by quiescent phases in between (Figure \ref{fig:fig4}a and Supplementary Figure 3). As for the 1:1 entrainment region, phase plane analysis revealed that the bursting behavior is caused by creating and destroying fixed points while driving the system back and forth the SNIC bifurcations. Due to the low forcing frequency, however, multiple cycles on the periodic orbit can be completed before the trajectory is recaptured by the emergence of the stable fixed point (Supplementary Video 4). This resembles the description of parabolic circle/circle bursting \cite{Izhikevich_2006}, except that the role of the slow variable driving the system back and forth the SNIC is played by the periodic forcing signal.

In case of the VdP$_{\textnormal{2g}}$ system, an unimodal distribution (with  decreasing ISI duration for increasing forcing frequency) is observed for most of the forcing interval, which is consistent with the large Arnold tongue for the 1:1 entrainment regime for this system (Figure \ref{fig:fig2}c, right). In agreement with the robust catch-and-release mechanism between two fixed points, no bursting is observed for the VdP$_{\textnormal{2g}}$ system.
For very high driving frequencies, however, the ISI distributions become disperse (inset in Figure \ref{fig:fig2}c, right) and the corresponding time-series reveal that the system exhibits complex behavior with quick divergence of the trajectories after small perturbations of the initial conditions (Figure \ref{fig:fig4}b), indicative of chaotic dynamics. Calculating the largest Lyapunov-exponents (LLEs) for all three systems shows that only the VdP$_{\textnormal{2g}}$ system exhibits chaotic behavior in the $(A,\omega)$-space considered here (Figure \ref{fig:fig4}c and Figure \ref{fig:fig2}a, cyan crosses). 
Together, these results demonstrate that the response properties of ghost cycles are markedly different from the dynamics of conventional slow-fast systems.

\subsection{Slow-fast and ghost cycle dynamics in the Morris-Lecar model}

To demonstrate that the observed response properties of ghost cycles and slow-fast systems are not peculiarities of the chosen model systems, we next validate the above obtained results for the Morris-Lecar (ML) model\cite{Morris_1981}, another typical slow-fast system that features both excitable and oscillatory dynamics. 
The ML system\cite{Tsumoto_2006} is given by:
\begin{align}\label{eq:MLsys}
\begin{split}
    C_m \dot{V} &= -g_L(V-V_L)-g_{Ca}M_{\infty}(V-V_{Ca})-g_K N (V-V_K) + (I_{ext} + A\ \sin(\omega t)) \\
    \dot{N} &= \frac{N_\infty-N}{\tau_N},
\end{split}
\end{align}

\noindent where $V$ is the membrane potential in mV, $N$ is the dimensionless activation variable and

\begin{align*}
    M_\infty &= \frac{1}{2} \left(1+\tanh\left(\frac{V-V_1}{V_2}\right)\right)\\
    N_\infty &= \frac{1}{2} \left(1+\tanh\left(\frac{V-V_3}{V_4}\right)\right)\\
    \tau_N &= 1/\left(\phi\cosh\left(\frac{V-V_3}{2V_4}\right)\right).
\end{align*}

\begin{figure}[h]
\includegraphics[width=\textwidth]{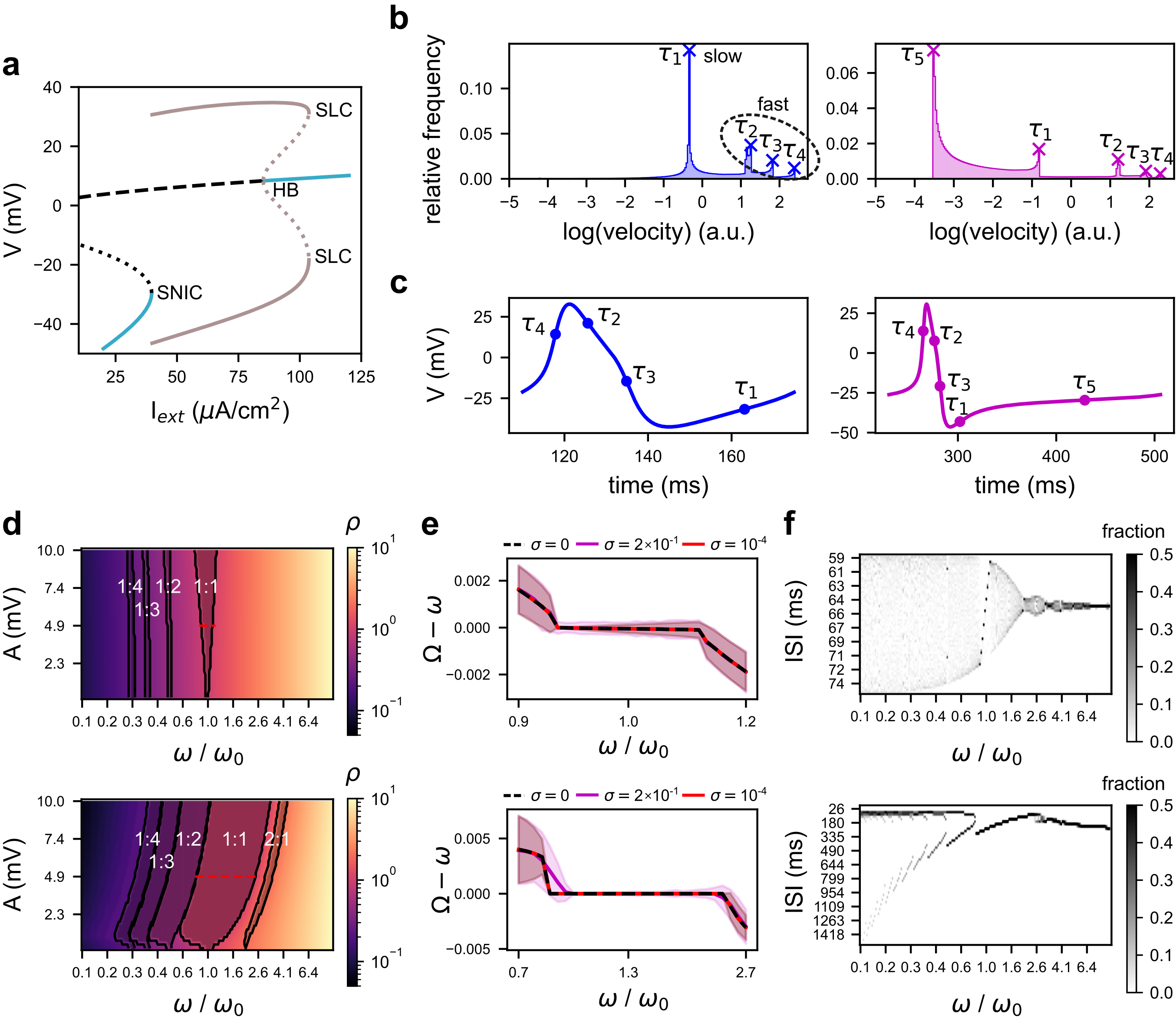}
\caption{Slow-fast and ghost-cycle regimes in the Morris-Lecar (ML) model. (a) Bifurcation diagram. Dashed black/gray line: unstable spiral/limit cycle; grey line: stable limit cycle; dotted black line: saddle; teal line: stable fixed point. SNIC bifurcation at $I_{ext} = 39.69345\ \mu A/cm^2$. (b) Histograms of the velocity distributions from system's trajectories in the slow-fast regime (left) at $I_{ext} = 55\ \mu A/cm^2$ and the 1-ghost cycle regime (right) at $I_{ext} = 40.19345\ \mu A/cm^2$. Indicated peaks are taken as representative time-scales of the systems. (c) Time-series of a single period of the ML system in the slow-fast regime (left) 1-ghost cycle regime (right). (d-f) Response of the ML system in the slow-fast (top row) and 1-ghost cycle regime (bottom row) to periodic forcing. (d) Corresponding Arnold tongues indicating areas of n:m entrainment as a function of the amplitude $A$ and frequency $\omega$ of the forcing (normalized to $\omega_0$). Color encodes the winding number $\rho = \frac{T_{out}}{T_{in}}$. (e) Difference of the observed frequency $\Omega$ and the forcing frequency $\omega$ at various noise intensities in the parameter range indicated by the dashed red lines in (d). Plateaus correspond to cross-sections through the 1:1 entrainment tongues. (f) Distribution of the interspike intervals (ISIs) as a function of the forcing frequency.}
\label{fig:fig5}
\end{figure}

\noindent As before, $A$ is the amplitude and $\omega$ the frequency of external periodic forcing. The ML system is parameterized to display type-I neuronal activity (cf. \cite{Morris_1981,Tsumoto_2006} for biophysical details):

\begin{align*}
&C_m = 20 \ (\mu F/cm^2), V_1 = -1.2 \ (mV), V_2 = 18 \ (mV), V_3 = 12 \ (mV), V_4 = 17 \ (mV), \\
 & g_L = 2 \ (mS/cm^2), g_K = 8 \ (mS/cm^2), g_{Ca} = 4 \ (mS/cm^2), \phi = 0.06667 \ (s^{-1}), \\
 &V_L = -60\ (mV), V_K = -80 \ (mV), V_{Ca} = 120 \ (mV).
\end{align*}
\noindent  In type-I neurons, oscillations are initiated via a single SNIC bifurcation as a function of an external input current $I_{ext}$ (Figure \ref{fig:fig5}a), thus allowing the emergence of a 1-ghost cycle when organized close to the SNIC. In this study we chose $I_{ext} = 55\ \mu A/cm^2$ for the slow-fast regime and $I_{ext} = 40.19345\ \mu A/cm^2$ for the 1-ghost cycle regime.

Numerical analysis of the time-scales of the systems shows that the type-I ML neurons feature several characteristic timescales corresponding to complex shape of the relaxation oscillations (Figure \ref{fig:fig5}b,c left). For simplicity, we consider $\tau_2-\tau_4$ to constitute the fast time-scale, since they define the shape of a single spike. As for the VdP$_{\text{1g}}$ system, however, the additional slow timescale of the 1-ghost cycle, that emerges for organization close to the SNIC bifurcation, dominates the dynamics of the ML oscillator (Figure \ref{fig:fig5}b,c right).

While the slow-fast regime of the type-I ML neuron shows an almost constant period with little variability for the full range of $\sigma$-values, the dependency of the ML period on noise intensity when operating in the 1-ghost cycle regime are similar to the observations in the 
VdP$_{\textnormal{1g}}$ model (Supplementary Figure 5). Moreover, in the periodically forced ML oscillator and its 1-ghost variant we find equivalent entrainment capabilities with robustness to noise, bursting behavior and changes in the ISI distributions as described for the VdP and VdP$_{\text{1g}}$ models (Figure \ref{fig:fig5}d-f, see also Supplementary Figures 6-7 for representative time-series). These results therefore demonstrate that the response properties of fast-slow and ghost cycle models are independent from the underlying model realization, but results from the mechanisms of emergence of the systems' dynamics.  

\section{\label{sec:conclusion} Discussion and conclusion}
Oscillatory processes with multiple time-scales which are prevalent in natural systems have generally been described as slow-fast systems, where the difference in time scales results from the presence of a small parameter ($\varepsilon$) in the system's equations. For the VdP system, there is a single normally hyperbolic equilibrium point at the origin, and for $\varepsilon>0$ sufficiently small there exists a slow normally hyperbolic critical manifold, such that the points on it can be viewed as hyperbolic points of the fast subsystem. Thus, stable and unstable branches of the critical manifold can be defined that govern the dynamics of the system. The period of the VdP oscillator is therefore not sensitive to noise, and the response to external forcing displays a typical Arnold tongue distribution.
In contrast, the dynamics of a ghost cycle such as the modified VdP$_{\text{2g}}$ system is governed predominantly by the trapping in the ghost states which determines the slow time-scale of the system, with rapid switching among them. The ghost states are weakly attracting sets that are not invariant \cite{Gorban_2013} and the response dynamics of the ghost cycles to external forcing are notably different, e.g. the systems entrain 1:1 to a broad range of external frequencies in a fashion that is robust to noise. 
Moreover, ghost cycles also exhibit additional dynamical responses to forcing, such as bursting. Interestingly, we have observed bursting behavior mainly for the one-ghost cycles of the VdP$_{\text{1g}}$ and ML systems.

These differences in the dynamics of slow-fast oscillators and ghost cycles can be understood from the non-autonomous description of the systems. Whereas the external forcing does not affect the qualitative organization of the phase space of the classical VdP system, the criticality (closeness to one or more SNIC bifurcations) through which ghost cycles generally emerge enables that even small amplitude forcing creates and destroys fixed points and ghost attractors, i.e. it continually reshapes the systems attractor landscape. The response behavior therefore depends also on the number and the particular phase space organization of the ghosts in the non-autonomous system (as exemplified by the release-and-catch mechanism between two stable fixed points that emerge and disappear in an alternating manner when the VdP$_{\text{2g}}$ system is periodically forced, underlying the ultra-large 1:1 Arnold tongue), as it depends on the frequency of the forcing that leads to the attractor landscape changes (resulting e.g. in 1:1 entrainment or to bursting for different frequencies in the VdP$_{\text{1g}}$ system). 

Recently proposed gene-regulatory network models that are characterized with three simultaneous SNIC bifurcations\cite{Jutras_Dub__2020, Farjami_2021,Yang_2022} give rise to a three-ghost cycle \cite{Koch_2023}, and models of recurrent neuronal networks\cite{Jordan_2021} suggest also the existence of cycles between four or more ghosts . Since the response of ghost cycles to periodic forcing depends on the orientation of the phase flow that is organized by the ghosts, it will be interesting to study the response properties if multi-ghost cycles to periodic forcing. Given that many of the one- and multi-ghost cycles seem to be typical for models of biological networks, we speculate that the flexibility (being able to exploit different dynamics via the changes in the attractor landscape) and robustness in the response to external forcing that emerges in these models could be a feature that biological networks evolved to exploit for information processing or biological computation. As ghost cycles emerge for system's organization at criticality (closeness to one or more SNIC bifurcations), biological systems  may exhibit feedback mechanisms for self-organization close to the critical point that enables them to utilize the computational capabilities of ghost cycles \cite{Durstewitz_2003,Stanoev_2020}. Future studies are thus required to test these hypotheses in the context of models for computational tasks in biological networks.

\begin{acknowledgments}
D.K. received funding by an EMBO Fellowship (Grant nr. ALTF 310-2021), A.K. acknowledges funding by the Lise Meitner Excellence Programme of the Max Planck Society.
\end{acknowledgments}

\providecommand{\noopsort}[1]{}\providecommand{\singleletter}[1]{#1}%

\section*{Supplementary Information}

\subsection*{Numerical methods}

\noindent Deterministic simulations were performed with a 4th order Runge-Kutta scheme using custom-made python code. Stochastic simulations were performed as described in \cite{Koch_2023} by modeling noise as a Wiener process where Gaussian white noise is introduced as an additive term at each time step, yielding a stochastic differential equation  in $It\hat{o}$ form, $\dot{X(t)} = f(X(t), t)dt + \sigma dW(t)$, where $dt$ denotes the step size, $\sigma(t)$ describes the additive noise and $W(t)$ denotes a Wiener process whose independent increments follow a normal distribution with $(\mu=0,SD=1)$ and amplitude $\sqrt{\Delta t}$. In both deterministic and stochastic simulations, a stepsize of $\Delta t = 0.05$ was used.

\noindent Bifurcation analysis was performed using the {\tt AUTO}-module of XPPAUT\cite{Ermentrout_2002} (\textcolor{blue}{\url{https://sites.pitt.edu/~phase/bard/bardware/xpp/xpp.html}}).

\noindent Largest Lyaponuv exponents were calculated by the method described in \cite{Benettin_1976} using custom-made python code.

\noindent Periods and interspike-intervals (ISIs) were determined from the peak-to-peak distance using the {\tt signal.find\_peaks} function from the python package SciPy  either directly from the the simulated timeseries (ML system) or its numerical derivative (VdP, VdP$_{\text{1g}}$ and VdP$_{\text{2g}}$ systems).

\noindent Timescales were determined numerically from phase space trajectories by identifying the peaks of their velocity distribution histograms and the time constants defining different timescales were calculated as $\tau_i = v_{p_i}^{-1}$, where $v_{p_i}$ denotes the velocity corresponding to the i-th peak of the histogram. 

\noindent To study the response to external forcing we calculated the winding number $\rho = \frac{T_{out}}{T_{in}}$, where $T_{out}$ is the observed period of the system subject to external forcing and $T_{in} = \frac{1}{\omega}$ is the period of the forcing signal, using forcing frequencies ranging from from $0.1$ to $10$-fold of the system's intrinsic frequency (i.e. in absence of external forcing) $\omega_0$, low amplitudes (up to $A=0.2$ for VdP, VdP$_{\text{1g}}$ and VdP$_{\text{1g}}$; up to $A=10\ \mu A / cm^2$ for ML). Arnold tongues corresponding to $n:m$ entrainment (restricted to $m,n \in \{1,2,3,4\}$ in this study, where $n$ periods of the forced oscillator can be observed for every $m$ periods of the external forcing signal) can be identified as areas in $(A,\omega)\subset\mathbb{R}^2$ where $\rho = \frac{m}{n}$, which was evaluated numerically via $|\rho - \frac{m}{n}| < 0.01$ using timeseries from a total simulation time of $t = 60\cdot T_0, T_0 = \frac{1}{\omega_0}$. Only Arnold tongues covering at least 1\% of the considered $(A,\omega)$ subspace are shown in this study. \\

\noindent All codes for reproducing the results and figures from this study will be available on GitHub upon final publication at: \textcolor{blue}{\url{https://github.com/KochLabCode/GhostCycles}}\\ \\
\noindent The versions of the utilized python packages were as follows:

\begin{center}
\begin{tabular}{|cc|}
    \hline
     \textbf{Package} & \textbf{Version} \\
     \hline
     Numpy & 1.16.6 \\
     SciPy &  1.7.1 \\
     Matplotlib &  3.4.3 \\
     \hline
\end{tabular}
\end{center}


\providecommand{\noopsort}[1]{}\providecommand{\singleletter}[1]{#1}%

\newpage 
\subsection*{ Supplementary Figures}
\renewcommand{\figurename}{Supplementary Fig.}
\counterwithin{figure}{section}
\renewcommand\thefigure{\arabic{figure}}
\setcounter{figure}{0} 
\vspace{1 cm}

\begin{figure}[h]
\begin{center}
   \includegraphics[width=.5\textwidth]{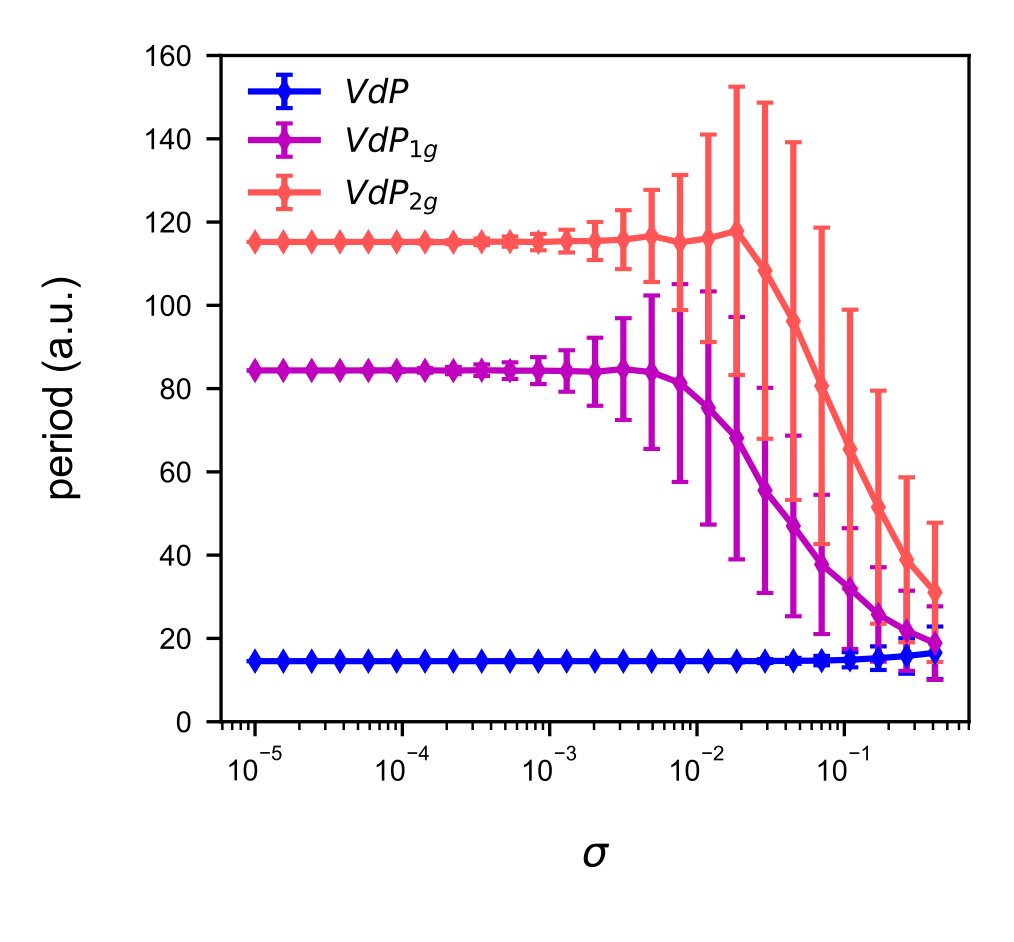}
\end{center}
\caption{Oscillation period as a function of the additive noise intensity $\sigma$ for the autonomous VdP, VdP$_\text{1g}$ and VdP$_\text{2g}$ systems. Data are shown as mean\textpm s.d. from 50 repetitions.}
\end{figure}

\begin{figure}[h]
\begin{center}
   \includegraphics[width=\textwidth]{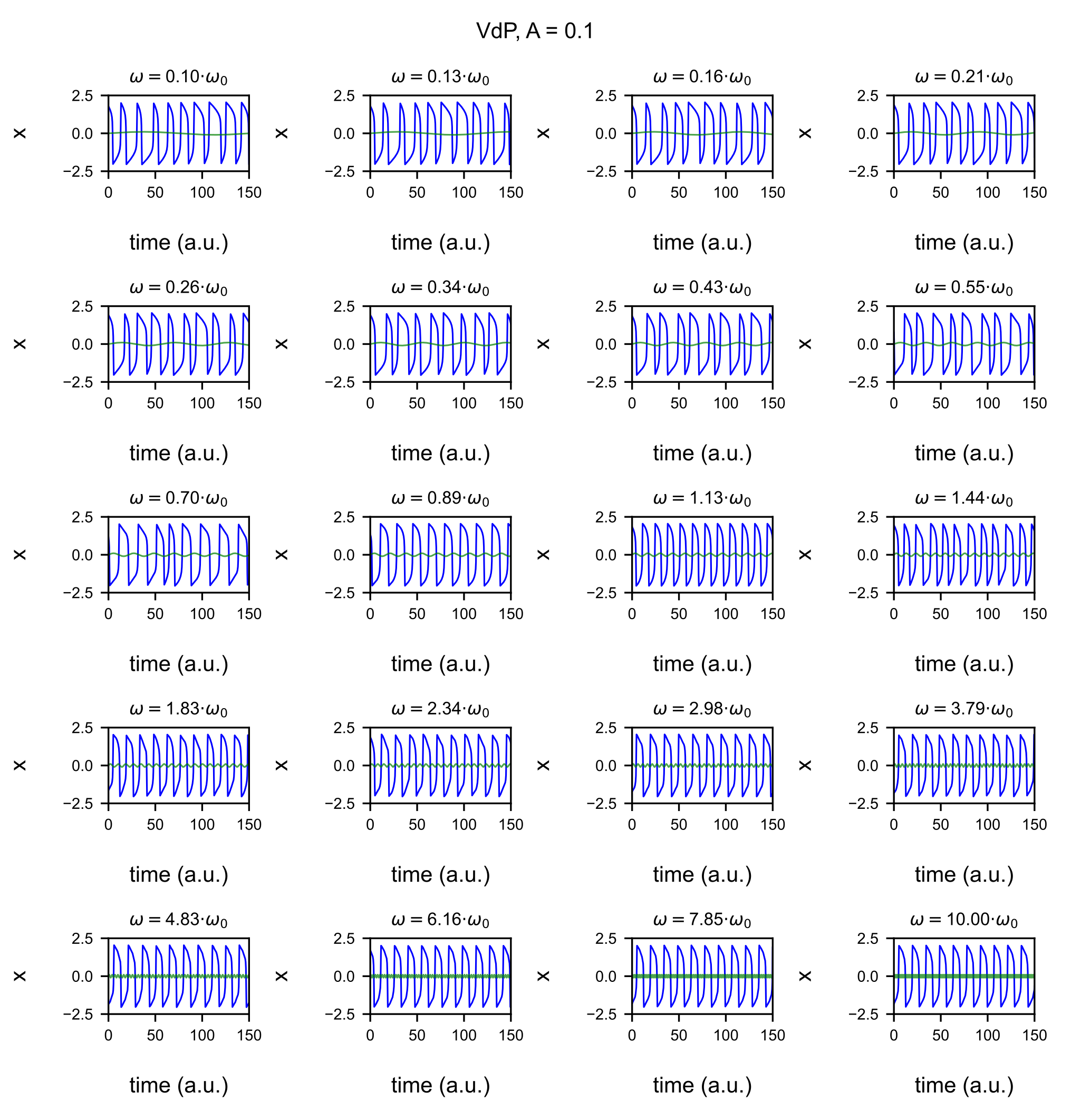}
\end{center}
\caption{Representative time courses from simulations of the forced VdP model selected from the full range of forcing frequencies $\omega$ at amplitude $A = 0.1$. Forcing input shown in green. }
\end{figure}

\begin{figure}[h]
\begin{center}
   \includegraphics[width=\textwidth]{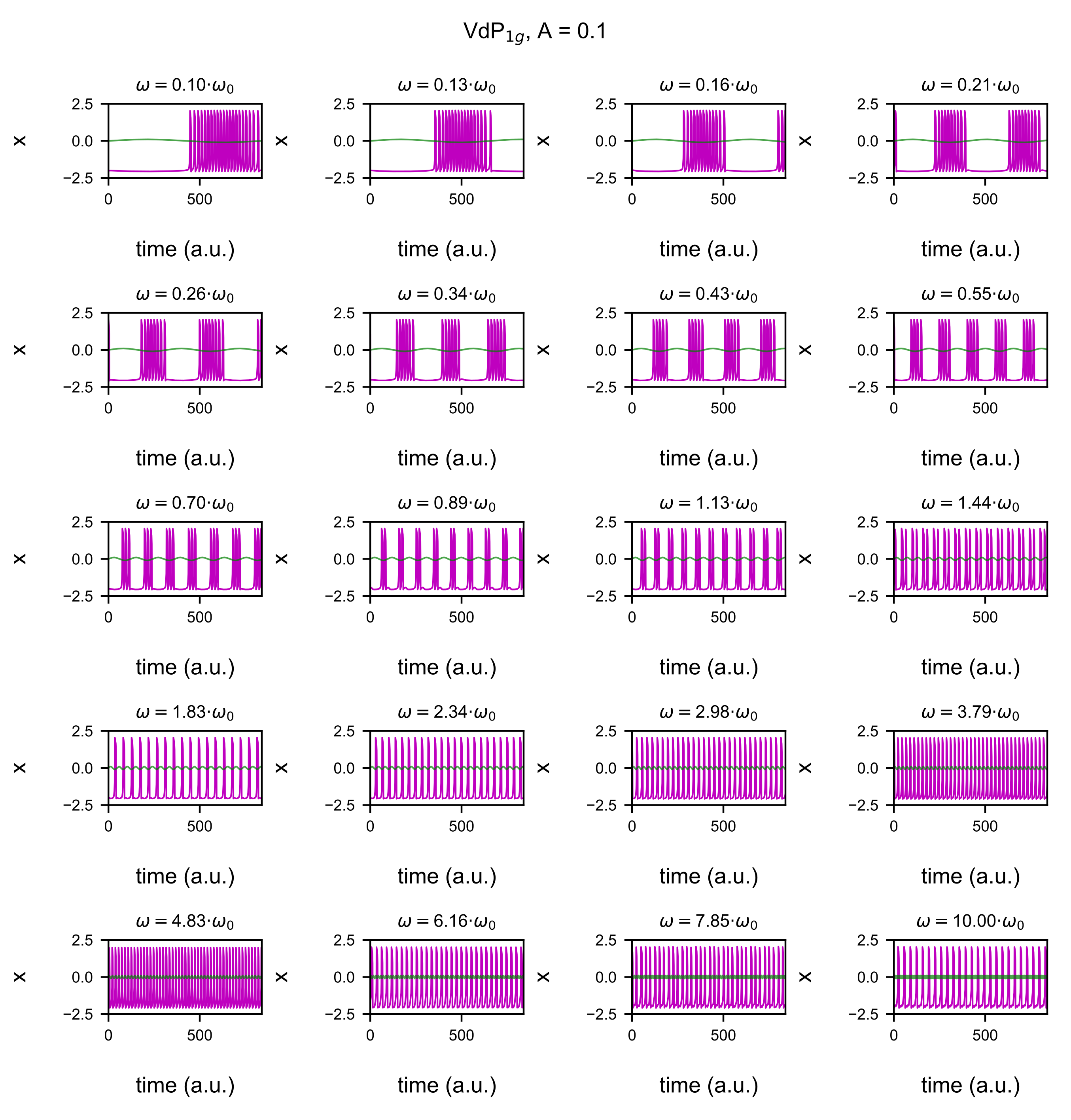}
\end{center}
\caption{Representative time courses from simulations of the forced VdP$_\text{1g}$ model selected from the full range of forcing frequencies $\omega$ at amplitude $A = 0.1$. Forcing input shown in green.}
\end{figure}

\begin{figure}[h]
\begin{center}
   \includegraphics[width=\textwidth]{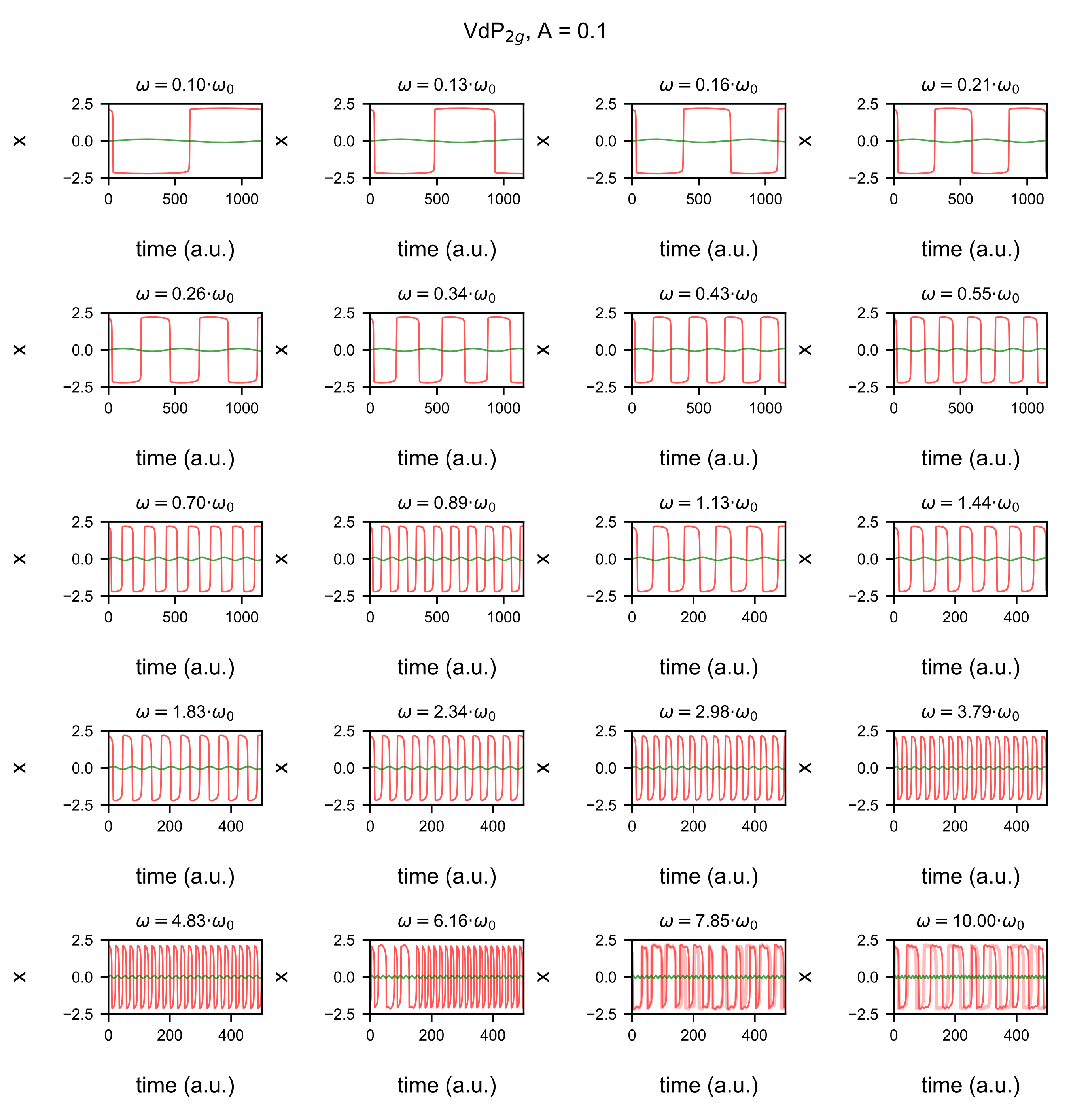}
\end{center}
\caption{Representative time courses from simulations of the forced VdP$_\text{2g}$ model selected from the full range of forcing frequencies $\omega$ at amplitude $A = 0.1$. Forcing input shown in green.}
\end{figure}

\begin{figure}[h]
\begin{center}
   \includegraphics[width=.5\textwidth]{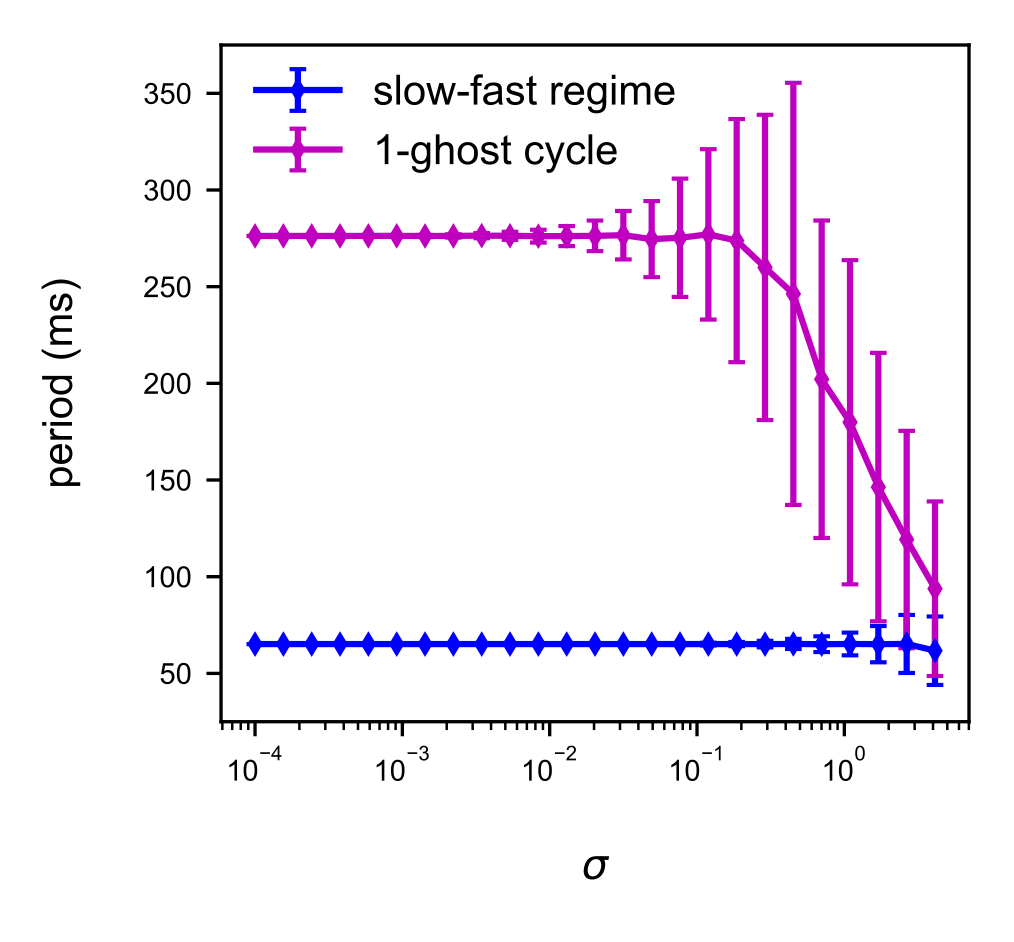}
\end{center}
\caption{Oscillation period of the autonomous ML system in the slow-fast or 1-ghost cycle regime as a function of the additive noise intensity $\sigma$. Data are shown as mean\textpm  s.d. from 50 repetitions.}
\end{figure}

\begin{figure}[h]
\begin{center}
   \includegraphics[width=\textwidth]{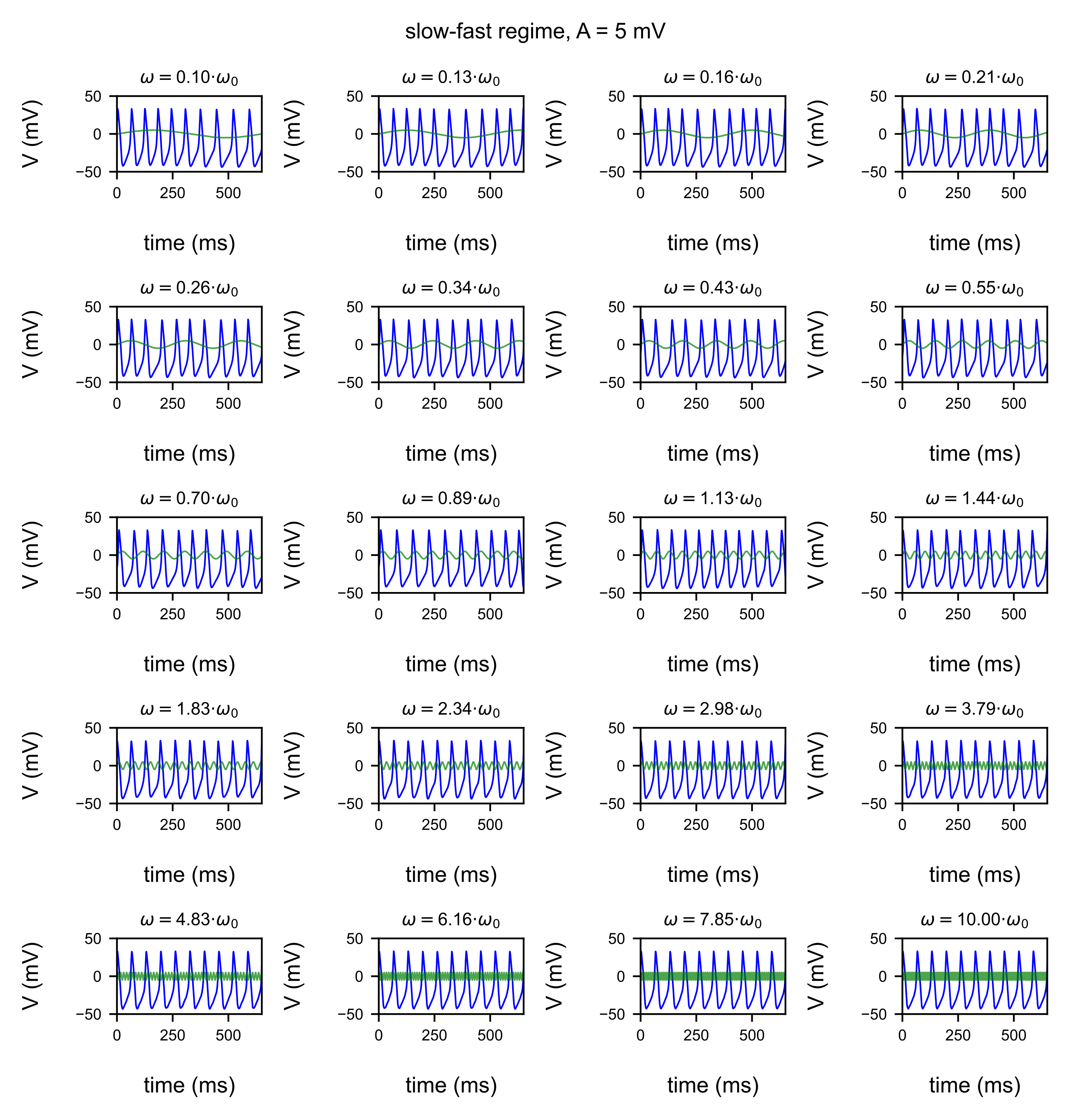}
\end{center}
\caption{Representative time courses from simulations of the forced ML model in the slow-fast regime selected from the full range of forcing frequencies $\omega$ at amplitude $A = 5\ mV$. Forcing input shown in green.}
\end{figure}

\begin{figure}[h]
\begin{center}
   \includegraphics[width=\textwidth]{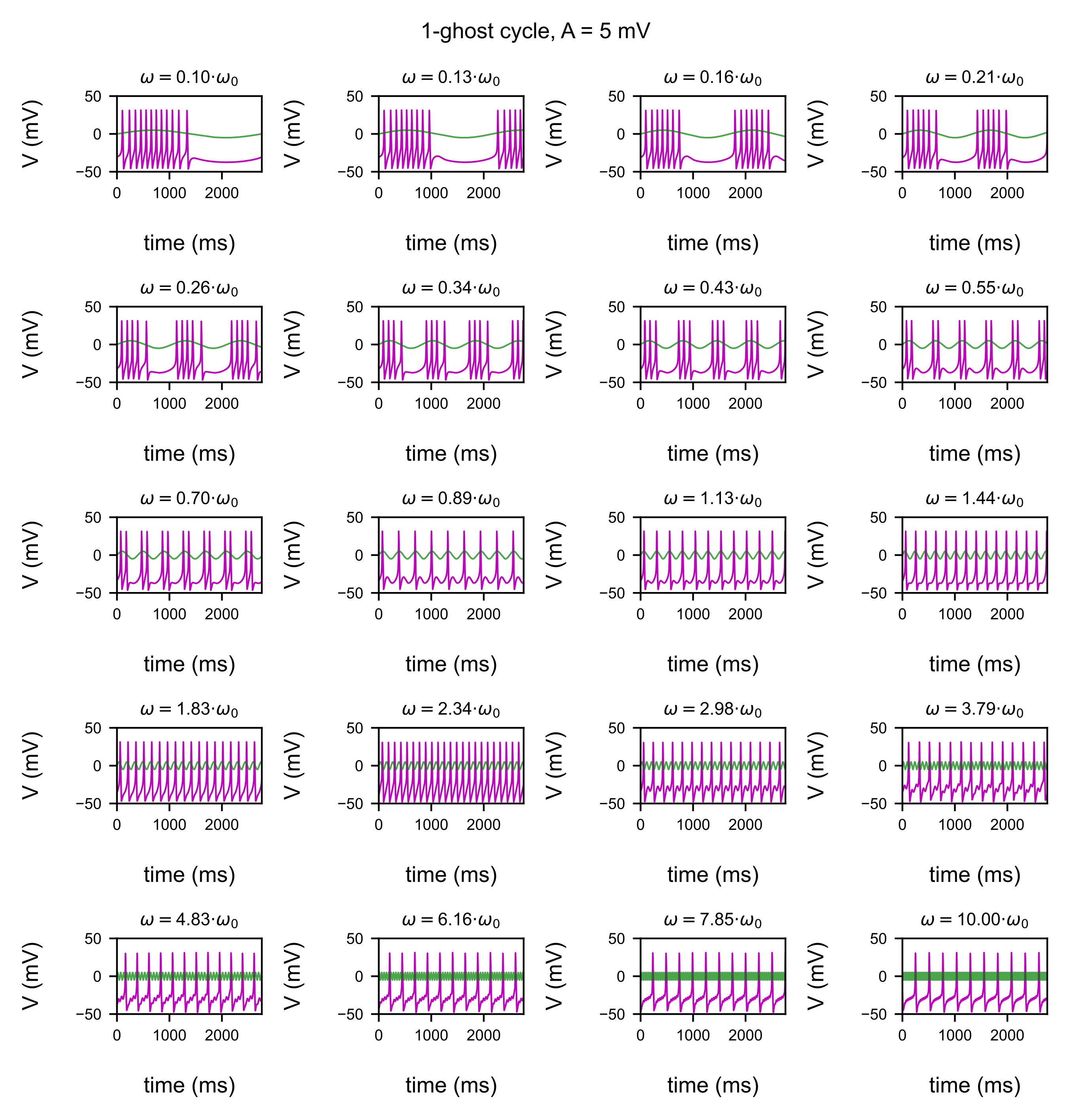}
\end{center}
\caption{Representative time courses from simulations of the forced ML model in the 1-ghost cycle regime selected from the full range of forcing frequencies $\omega$ at amplitude $A = 5\ mV$. Forcing input shown in green.}
\end{figure}

\end{document}